\documentclass[%
 reprint,
 superscriptaddress,
 amsmath,amssymb,
 aps,
 pre,
nofootinbib,
]{revtex4-2}

\usepackage[usenames,dvipsnames]{xcolor}

\usepackage{amsthm}
\usepackage{amssymb}

\usepackage{diagbox}

\usepackage{graphicx}
\usepackage{dcolumn}
\usepackage{bm}
\usepackage[unicode, colorlinks, urlcolor=blue, linkcolor=blue, pagecolor=blue, citecolor=blue]{hyperref} 
\usepackage[T2A]{fontenc}
\usepackage[utf8]{inputenc}
\usepackage[russian,english]{babel}
\usepackage{comment}
\addto\captionsenglish{}
\usepackage{bigstrut}
\usepackage{float}

\let\oldcfrac\cfrac
\renewcommand{\cfrac}[2]{\oldcfrac{#1}{#2}\,}
\usepackage{comment}
\usepackage{hhline}
\begin{document}

\title{One--Component Plasma Equation of State Revisited\\ via Angular--Averaged Ewald Potential}

\author{G. S. Demyanov}
\affiliation{Joint Institute for High Temperatures, Izhorskaya 13 Bldg 2, Moscow 125412, Russia}
\affiliation{Moscow Institute of Physics and Technology, Institutskiy Pereulok 9, Dolgoprudny, Moscow Region, 141701, Russia}
\author{P. R. Levashov}
\affiliation{Joint Institute for High Temperatures, Izhorskaya 13 Bldg 2, Moscow 125412, Russia}
\affiliation{Moscow Institute of Physics and Technology, Institutskiy Pereulok 9, Dolgoprudny, Moscow Region, 141701, Russia}

\date{\today}

\begin{abstract}
	We present analytic fits of classical one--component plasma (OCP) internal energy over a wide range of coupling parameter $0.01\le\Gamma\le170$ using Monte--Carlo data in the thermodynamic limit. We extend the dataset obtained in \href{https://doi.org/10.1103/PhysRevE.106.015204}{[Demyanov and Levashov, Phys. Rev. E \textbf{106}, 015204 (2022)]} using the angular--averaged Ewald potential with additional points at strong coupling ($\Gamma=120,150,170$). We then fit two frequently used functional forms for the OCP equation of state: (i) a five-parameter equation by Caillol \href{https://doi.org/10.1063/1.479965}{[J. Chem. Phys. 111, 6538–6547 (1999)]}, and (ii) the equation by Potekhin and Chabrier \href{https://doi.org/10.1103/PhysRevE.62.8554}{[Phys. Rev. E \textbf{62}, 8554 (2000)]} that enforces the Debye--H\"uckel limit. The presented fits reproduce our MC data within statistical uncertainties, recovering the correct weak-coupling behavior. Coefficients, recommended validity ranges, and comparisons to prior analytical and simulation results are provided.
\end{abstract}

\maketitle

In our previous work~\cite{Demyanov:PRE:2022} we computed a tabulated OCP equation of state (EOS)  over $0.01\leq \Gamma \leq 100$ using up to $10^6$ particles in Monte--Carlo (MC) simulations. This accuracy became enabled via the angular-averaged Ewald potential (AAEP) that effectively captures the long-range Coulomb interactions. Our results were compared with the MC simulations~\cite{Caillol:1999,Caillol:2010}, hypernetted chain (HNC) integral equation solution~\cite{Caillol:2010}, and cluster expansion by Ortner~\cite{Ortner:1999}. In contrast to Caillol and Gilles~\cite{Caillol:2010}, we found excellent agreement with the analytical result at $\Gamma\leq 0.1$. While our prior work tabulated the internal energy, a compact EOS fit across the full fluid regime was not provided. This note supplies such an EOS and compares two classes of interpolants, clarifying their asymptotics and range of validity.  We also extend our results to the strong coupling regime by adding MC simulation data at $\Gamma=120,150,170$. We use the same notations as in Ref.~\cite{Demyanov:PRE:2022}.

We employ a standard MC simulation technique for the classical OCP with the AAEP (see Ref.~\cite{Demyanov:PRE:2022} for details). Simulations are initialized from random ion configurations and equilibrated before the main simulation section. After equilibration, we perform $m_{\text{tot}}=10^7$ MC steps for $N=10^2,10^3,10^4,10^5,10^6$ at $\Gamma=0.01,0.05,0.1,1$. To reduce statistical errors at strong coupling, for $\Gamma=10$--$170$ we use $m_{\text{tot}}=10^8$ steps for $N$ in the range $10^2$--$10^5$.
Statistical uncertainties are estimated via the standard block averaging~\cite[Sec.~11.4]{GouldHarvey1996Aitc} with $n_b$ blocks; each block contains $2\times10^6$ energy samples. 

 \begin{table}[h!]
	\centering
	\caption{Parameters of MC simulations with the AAEP.}
	\begin{tabular}{|c|c|c|c|}
		\hline
		\multicolumn{1}{|c|}{$\Gamma$} & $N$ & $m_{tot}$ & \multicolumn{1}{c|}{$n_b$} \bigstrut\\
		\hline
		$0.01$ & $10^2$, $10^3$, $10^4$, $5\times 10^4$, $10^5$, $10^6$ & $10^7$ & $5$ \bigstrut\\
		\hline
		$0.05$ & $10^2$, $10^3$, $10^4$, $10^5$, $10^6$ & $10^7$ & $5$ \bigstrut\\
		\hline
		$0.1$ & $10^2$, $10^3$, $10^4$, $10^5$, $10^6$ & $10^7$ & $5$ \bigstrut\\
		\hline
		$1$ & $10^2$, $10^3$, $10^4$, $10^5$, $10^6$ & $10^7$ & $5$ \bigstrut\\
		\hline
		$10$ & $10^2$, $10^3$, $10^4$, $10^5$ & $10^8$ & $50$ \bigstrut\\
		\hline
		$100$ & $10^2$, $150$, $10^3$, $10^4$, $10^5$ & $10^8$ & $50$ \bigstrut\\
		\hline
		
		$120$ & $10^3$, $5\times 10^3$, $10^4$, $5\times 10^4$, $10^5$ & $10^8$ & $50$ \bigstrut\\
		\hline
		$150$ & $10^3$, $5\times 10^3$, $10^4$, $5\times 10^4$, $10^5$ & $10^8$ & $50$ \bigstrut\\
		\hline
		$170$ & $10^3$, $5\times 10^3$, $10^4$, $5\times 10^4$, $10^5$ & $10^8$ & $50$ \bigstrut\\
		\hline
	\end{tabular}%
	\label{tab:techparams}%
\end{table}%

All simulation parameters extended relative to Ref.~\cite{Demyanov:PRE:2022} are collected in Table~\ref{tab:techparams}, and raw MC results for $-U/N$ as functions of $N$ are summarized in Table~\ref{tab:resNaaep}.

\begin{table*}[ht!]
	\centering
	\caption{MC results for $-U/N$ at $\Gamma = 0.1, 1, 10, 100$ as a function of $N$ using the AAEP. The digits in brackets correspond to one standard deviation. 
	}
	\label{tab:addlabel1}
	\begin{tabular}{|c|cccccccc|} 
		\hline
		\diagbox{$\Gamma$}{$N$}                     & $10^2$       & $150$        & $10^3$        & $5\times 10^3$ & $10^4$        & $5\times 10^4$   & $10^5$        & $10^6$         \\ 
		\hline
		$0.01$                                      & 0.0020724(45)& ---           & 0.0011814(96) & ---            & 0.000917(12) & 0.000875(11)  & 0.000866(15) & 0.000863(14)  \bigstrut[t]\\
		$0.05$                                      & 0.012662(30) & ---           & 0.009855(38)  & ---            & 0.009439(31) & ---  &  0.009405(54)&  0.009409(31)\bigstrut[t]\\
		$0.1$                                      & 0.029951(50) & ---           & 0.026144(11)  & ---            & 0.025756(63) & ---  & 0.025704(21) & 0.025691(37)  \bigstrut[t]\\
		$1$                                        & 0.57682(21)  & ---           & 0.57201(15)   & ---            & 0.57144(21)  & ---  & 0.57149(34)  & 0.57142(14)  \bigstrut[t]\\ 
		$10$                                       & 8.0043(14)   & ---           & 7.9985(13)    & ---            & 7.9982(14)   &  --- & 7.9982(14)   & ---            \bigstrut[t]\\  
		$100$                                      & 87.485(11)   & 87.645(11)    & 87.5295(62)   & ---            & 87.5248(50)  & ---  & 87.5235(33)  & ---             \bigstrut[b]\\ 
		$120$                                      & ---          & ---           & 105.3489(86)  & 105.3468(82)   & 105.3463(54) & 105.3460(37)  & 105.3461(24) &  ---           \bigstrut[b]\\ 
		$150$                                      & ---          & ---           & 132.111(10) & 132.1122(93)   & 132.1124(71) & 132.1091(33)  & 132.1104(24) & ---             \bigstrut[b]\\ 
		$170$                                      & ---          & ---           & 149.969(13) & 149.9727(90)   & 149.9743(81) & 149.9669(44)  & 149.9697(34) & ---             \bigstrut[b]\\ 
		\hline
	\end{tabular}
	\label{tab:resNaaep}
\end{table*}

To obtain the thermodynamic limit $N\to\infty$, we fit the $N$-dependence of energy per particle using the following function:
\begin{equation}
	\label{eq:fitFunc}
	\frac{U}{N}\,(1/N) = \left(\frac{U}{N}\right)_\infty + b\left(\frac{1}{N}\right)^{\gamma}.
\end{equation}
Here $\left(U/N\right)_\infty$, $b$, and $\gamma$ are fit parameters determined \emph{for each} $\Gamma$. For large $\Gamma\geq 120$ where only $N\ge10^3$ are used, we set $\gamma=1$ (linear extrapolation in $1/N$). The resulting thermodynamic limit values are listed in Table~\ref{tab:addlabel} with the MC data by Caillol \emph{et al.}~\cite{Caillol:1999,Caillol:2010}, the cluster expansion~\cite{Ortner:1999}, Debye--H\"uckel asymptotic, and HNC results from Ref.~\cite{Caillol:2010}.

\begin{table*}[ht!]
	\centering
	\caption{Thermodynamic limit of MC results for $-U/(N\Gamma)$. 
		The results obtained using the AAEP agree very well with Caillol \emph{et al.} at $\Gamma = 1$, 
		and with the theoretical values at $\Gamma = 0.01$--$0.1$. 
		The digits in brackets correspond to one standard deviation,~$\sigma$. 
		The coefficients for EOS I and EOS II can be found in the last column of Tables~\ref{tab:EOS_OCP} and \ref{tab:EOS_OCP_II}, respectively. 
		Additional columns $z_I$ and $z_{II}$ show the normalized deviations $z_{I,II}=(U^{I,II}/(N\Gamma)-U^\text{MC (AAEP)}/(N\Gamma))/\sigma$, which are less than 1.}
	\begin{tabular}{|c|lllll|ll|ll|}
		\hline
		$\Gamma$ & \multicolumn{1}{c}{MC (AAEP)} 
		& \multicolumn{1}{c}{MC (Caillol \emph{et al.})} 
		& \multicolumn{1}{c}{Ortner} 
		& \multicolumn{1}{c}{DH} 
		& \multicolumn{1}{c|}{HNC} 
		& \multicolumn{1}{c}{EOS I} 
		& \multicolumn{1}{c|}{$z_I$} 
		& \multicolumn{1}{c}{EOS II} 
		& \multicolumn{1}{c|}{$z_{II}$} \\
		\hline
		0.01 & 0.08611(42)\footnote{Only $N\geq 10^4$ were used for fitting \eqref{eq:fitFunc}} & \multicolumn{1}{c}{---} & 0.086193 & 0.086603 & 0.086198\footnote{Obtained from HNC fit \cite[Tab. 1]{Caillol:2010} \label{note1}} & \multicolumn{1}{c}{---} & \multicolumn{1}{c|}{---} & 0.08641216 & +0.72 \\
		0.05 & 0.18774(30) & \multicolumn{1}{c}{---} & 0.187739 & 0.193649 & 0.187771$^{\text{\ref{note1}}}$
		& \multicolumn{1}{c}{---} & \multicolumn{1}{c|}{---} & 0.18803875 & +0.996 \\
		0.1 & 0.256975(35) & 0.25127(34) & 0.256992 & 0.273861 & 0.256885
		& 0.25697462 & –0.01 & 0.25696102 & –0.40 \\
		1 & 0.571414(24) & 0.571403(22) & 1.665188 & 0.866025 & 0.570455
		& 0.57141424 & +0.01 & 0.57141496 & +0.04 \\
		10 & 0.7998170(16) & 0.7997974(45) & \multicolumn{1}{c}{---} & \multicolumn{1}{c}{---} & \multicolumn{1}{c|}{---}
		& 0.79981703 & +0.02 & 0.79981702 & +0.01 \\
		100 & 0.8752382(55) & 0.8752693(24) & \multicolumn{1}{c}{---} & \multicolumn{1}{c}{---} & \multicolumn{1}{c|}{---} 
		& 0.87523855 & +0.06 & 0.87524317 & +0.90 \\
		120 & 0.8778840(4) & 0.8779087(30) & \multicolumn{1}{c}{---} & \multicolumn{1}{c}{---} & \multicolumn{1}{c|}{---}
		& 0.87788400 & 0.00 & 0.87788396 & –0.10 \\
		150 & 0.8807343(36) & 0.8807609(36) & \multicolumn{1}{c}{---} & \multicolumn{1}{c}{---} & \multicolumn{1}{c|}{---} 
		& 0.88073505 & +0.21 & 0.88073589 & +0.44 \\
		170 & 0.882172(8) & 0.8821895(30) & \multicolumn{1}{c}{---} & \multicolumn{1}{c}{---} & \multicolumn{1}{c|}{---}
		& 0.88216639 & –0.70 & 0.88217216 & +0.02 \\
		\hline
	\end{tabular}
	\label{tab:addlabel}%
\end{table*}%

Caillol~\cite{Caillol:1999} proposed the following five-parameter representation for the potential energy in the thermodynamic limit:
\begin{equation}
	\label{eq:eosI}
	\frac{\ U^\text{I}}{N} = A + B\,\Gamma + C\,\Gamma^{s} + D\,\Gamma^{-s},
\end{equation}
which generalizes Eq.~(5) of Ref.~\cite{Slattery:1980} (see also Eq.~(3) in Ref.~\cite{Stringfellow:1990}). The original coefficients and stated EOS~\eqref{eq:eosI} validity range $\Gamma\in[3,190]$ are reproduced in Table~\ref{tab:EOS_OCP}.

Despite its broad range, Eq.~\eqref{eq:eosI} has two shortcomings. First, the OCP melts around $\Gamma_m\simeq175$~\cite{Potekhin:2000}. Therefore, a single smooth expression cannot reflect the behavior of the liquid and crystalline phases at the melting point. Second, at small $\Gamma$ Eq.~\eqref{eq:eosI} fails to reproduce the correct Debye--H\"uckel behavior and diverge as $\Gamma\to 0$. As a result, the interval $0.1\leq\Gamma\leq 3$ remains poorly described by the original fit~\cite{Caillol:1999}. Also, it is probable that the authors of \cite{Caillol:1999} failed to describe weak coupling regime because the used MC method  apparently contained an error that led to a discrepancy between the analytical and calculated results at $\Gamma = 0.1$ (see Fig. 4(a) in Ref.~\cite{Demyanov:PRE:2022}).

\begin{figure*}[ht!]
	\centering
	\includegraphics[width=1\linewidth]{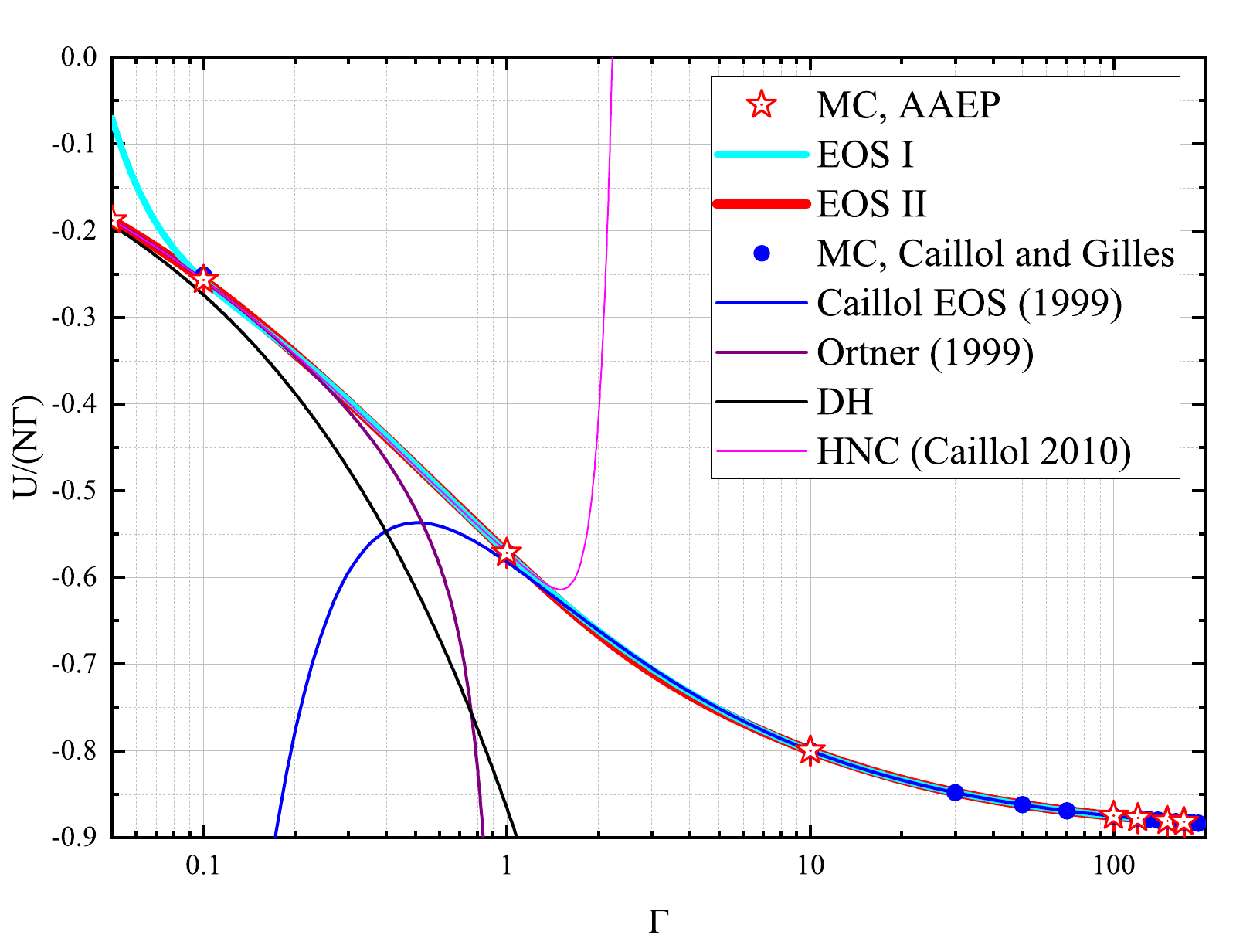}
	\caption{Comparison of the OCP potential energy, $U/(N\Gamma)$, dependence on the coupling parameter $\Gamma$. The red stars represent AAEP MC results, the cyan solid line and red solid line are EOS I~\eqref{eq:eosI} and EOS II~\eqref{eq:eosII}, respectively, fit to AAEP data. MC data by Caillol \textit{et al.}~\cite{Caillol:1999, Caillol:2010} are shown by the blue points, the blue solid line represents Caillol EOS~\eqref{eq:eosI}. The purple line shows Ortner's weak‑coupling expansion~\cite{Ortner:1999}, the black line shows the Debye–H\"uckel asymptote, and the magenta line shows the fit of HNC data from~\cite{Caillol:2010}. EOS II reproduces the Debye–H\"uckel (DH) limit as $ \Gamma\!\to\!0 $ and tracks HNC over $0.1\!\le\!\Gamma\!\le\!1 $, while both EOS curves agree with MC at strong coupling up to $ \Gamma=170 $. EOS I and Caillol EOS diverge at small $ \Gamma $.
	}
	\label{fig:neweospreprint}
\end{figure*}

Fitting Eq.~\eqref{eq:eosI} to our AAEP MC data (Table~\ref{tab:addlabel}) yields the coefficients listed in the second column of Table~\ref{tab:EOS_OCP}, and the corresponding curve is shown as the cyan solid line in Fig.~\ref{fig:neweospreprint}. Our fit corrects EOS behavior in the range $0.1\leq~\Gamma~\leq3$. However, when $\Gamma \leq 0.1$, our curve tends toward plus infinity, thus by definition it still does not capture the Debye--H\"uckel limit. Note that this equation incorrectly describes the HNC data in the range $0.1 \leq~\Gamma~\leq 1$, which is also due to the incorrect behavior of this curve.

To recover the correct weak-coupling asymptotic behavior, we use the EOS expression by Potekhin and Chabrier~\cite{Potekhin:2000}:
\begin{equation}
	\label{eq:eosII}
	\frac{U^\text{II}_\text{OCP}}{N}
	= \frac{A_1\,\Gamma^{3/2}}{\sqrt{\Gamma+A_2}}
	+ \frac{A_3\,\Gamma^{3/2}}{\Gamma+1}
	+ \frac{B_1\,\Gamma^{2}}{\Gamma+B_2}
	+ \frac{B_3\,\Gamma^{2}}{\Gamma^{2}+B_4},
\end{equation}
where $A_3=-\sqrt{3}/2 - A_1/\sqrt{A_2}$. Coefficients obtained by Potekhin and Chabrier for the Caillol data, and our new coefficients fitted to the AAEP MC data are provided in Table~\ref{tab:EOS_OCP_II}. The red solid curve in Fig.~\ref{fig:neweospreprint} demonstrates that EOS II reproduces the $\Gamma\ll 1$ limit, agrees with HNC results over $0.1\le\Gamma\le1$, and fits the AAEP MC data in strong coupling regime.

Note that both EOS I and EOS II reproduce the AAEP MC thermodynamic limit data within the obtained uncertainties across the entire fitted range.

\begin{table}[h!]
	\centering
	\caption{Coefficients for the OCP EOS~\eqref{eq:eosI}. The first column lists the coefficients reported in Ref.~\cite{Caillol:1999}. The second column contains the coefficients of the new EOS that correctly behaves in the range $\Gamma \in [0.1, 3]$ and is based upon AAEP MC data. The OCP EOS~\cite{Caillol:1999} is valid only at $\Gamma \in [3, 190]$.}
	\label{tab:EOS_OCP}
	\begin{tabular}{|c|c|c|}
		\hline
		& Caillol~\cite{Caillol:1999} & EOS I (AAEP data)  \\
		\hline
		$A$ & $-0.074970642$ & $-0.5425760160$  \\
		\hline
		$B$ & $-0.899588379$ & $-0.8987900948$  \\
		\hline
		$C$ & $0.494646173$ & $0.7585195383$  \\
		\hline
		$D$ & $-0.102192495$ & $0.1114323301$ \\
		\hline
		$s$ & $0.354161214$ & $0.2888260309$ \\
		\hline
		$\Gamma$ & $[3, 190]$ & $[0.1, 170]$  \\
		\hline
	\end{tabular}
\end{table}

\begin{table}[h!]
	\centering
	\caption{Coefficients for the OCP EOS~\eqref{eq:eosII}. 
		The first column lists coefficients reported in Ref.~\cite{Potekhin:2000} based on Caillol data~\cite{Caillol:1999}, the second is based on our AAEP MC data.}
	\label{tab:EOS_OCP_II}
	\begin{tabular}{|c|c|c|}
		\hline
		 & Caillol~\cite{Caillol:1999, Potekhin:2000} & EOS II (AAEP data)  \\
		\hline
		$A_1$ & -0.907347 &  -0.7678971255 \\
		\hline
		$A_2$ & 0.62849 &  0.5059769059 \\
		\hline
		$B_1$ &  $4.50\times 10^{-3}$ & -0.1336099542 \\
		\hline
		$B_2$ & 170 &  3.588111959\\
		\hline
		$B_3$ & $-8.4\times 10^{-5}$ & -0.1440005434 \\
		\hline
		$B_4$ & $3.70\times 10^{-3}$ & 5.177602644 \\
		\hline
	\end{tabular}
\end{table}


\begin{thebibliography}{8}%
\makeatletter
\providecommand \@ifxundefined [1]{%
 \@ifx{#1\undefined}
}%
\providecommand \@ifnum [1]{%
 \ifnum #1\expandafter \@firstoftwo
 \else \expandafter \@secondoftwo
 \fi
}%
\providecommand \@ifx [1]{%
 \ifx #1\expandafter \@firstoftwo
 \else \expandafter \@secondoftwo
 \fi
}%
\providecommand \natexlab [1]{#1}%
\providecommand \enquote  [1]{``#1''}%
\providecommand \bibnamefont  [1]{#1}%
\providecommand \bibfnamefont [1]{#1}%
\providecommand \citenamefont [1]{#1}%
\providecommand \href@noop [0]{\@secondoftwo}%
\providecommand \href [0]{\begingroup \@sanitize@url \@href}%
\providecommand \@href[1]{\@@startlink{#1}\@@href}%
\providecommand \@@href[1]{\endgroup#1\@@endlink}%
\providecommand \@sanitize@url [0]{\catcode `\\12\catcode `\$12\catcode
  `\&12\catcode `\#12\catcode `\^12\catcode `\_12\catcode `\%12\relax}%
\providecommand \@@startlink[1]{}%
\providecommand \@@endlink[0]{}%
\providecommand \url  [0]{\begingroup\@sanitize@url \@url }%
\providecommand \@url [1]{\endgroup\@href {#1}{\urlprefix }}%
\providecommand \urlprefix  [0]{URL }%
\providecommand \Eprint [0]{\href }%
\providecommand \doibase [0]{https://doi.org/}%
\providecommand \selectlanguage [0]{\@gobble}%
\providecommand \bibinfo  [0]{\@secondoftwo}%
\providecommand \bibfield  [0]{\@secondoftwo}%
\providecommand \translation [1]{[#1]}%
\providecommand \BibitemOpen [0]{}%
\providecommand \bibitemStop [0]{}%
\providecommand \bibitemNoStop [0]{.\EOS\space}%
\providecommand \EOS [0]{\spacefactor3000\relax}%
\providecommand \BibitemShut  [1]{\csname bibitem#1\endcsname}%
\let\auto@bib@innerbib\@empty
\bibitem [{\citenamefont {Demyanov}\ and\ \citenamefont
  {Levashov}(2022)}]{Demyanov:PRE:2022}%
  \BibitemOpen
  \bibfield  {author} {\bibinfo {author} {\bibfnamefont {G.~S.}\ \bibnamefont
  {Demyanov}}\ and\ \bibinfo {author} {\bibfnamefont {P.~R.}\ \bibnamefont
  {Levashov}},\ }\bibfield  {title} {\bibinfo {title} {One-component plasma of
  a million particles via angular-averaged ewald potential: A monte carlo
  study},\ }\href {https://doi.org/10.1103/PhysRevE.106.015204} {\bibfield
  {journal} {\bibinfo  {journal} {Phys. Rev. E}\ }\textbf {\bibinfo {volume}
  {106}},\ \bibinfo {pages} {015204} (\bibinfo {year} {2022})}\BibitemShut
  {NoStop}%
\bibitem [{\citenamefont {Caillol}(1999)}]{Caillol:1999}%
  \BibitemOpen
  \bibfield  {author} {\bibinfo {author} {\bibfnamefont {J.~M.}\ \bibnamefont
  {Caillol}},\ }\bibfield  {title} {\bibinfo {title} {{Thermodynamic limit of
  the excess internal energy of the fluid phase of a one-component plasma: A
  Monte Carlo study}},\ }\href {https://doi.org/10.1063/1.479965} {\bibfield
  {journal} {\bibinfo  {journal} {The Journal of Chemical Physics}\ }\textbf
  {\bibinfo {volume} {111}},\ \bibinfo {pages} {6538} (\bibinfo {year}
  {1999})}\BibitemShut {NoStop}%
\bibitem [{\citenamefont {Caillol}\ and\ \citenamefont
  {Gilles}(2010)}]{Caillol:2010}%
  \BibitemOpen
  \bibfield  {author} {\bibinfo {author} {\bibfnamefont {J.-M.}\ \bibnamefont
  {Caillol}}\ and\ \bibinfo {author} {\bibfnamefont {D.}~\bibnamefont
  {Gilles}},\ }\bibfield  {title} {\bibinfo {title} {{An accurate equation of
  state for the one-component plasma in the low coupling regime}},\ }\href
  {https://doi.org/10.1088/1751-8113/43/10/105501} {\bibfield  {journal}
  {\bibinfo  {journal} {Journal of Physics A: Mathematical and Theoretical}\
  }\textbf {\bibinfo {volume} {43}},\ \bibinfo {pages} {105501} (\bibinfo
  {year} {2010})}\BibitemShut {NoStop}%
\bibitem [{\citenamefont {Ortner}(1999)}]{Ortner:1999}%
  \BibitemOpen
  \bibfield  {author} {\bibinfo {author} {\bibfnamefont {J.}~\bibnamefont
  {Ortner}},\ }\bibfield  {title} {\bibinfo {title} {{Equation of states for
  classical Coulomb systems: Use of the Hubbard-Schofield approach}},\ }\href
  {https://doi.org/10.1103/PhysRevE.59.6312} {\bibfield  {journal} {\bibinfo
  {journal} {Phys. Rev. E}\ }\textbf {\bibinfo {volume} {59}},\ \bibinfo
  {pages} {6312} (\bibinfo {year} {1999})}\BibitemShut {NoStop}%
\bibitem [{\citenamefont {Gould}\ and\ \citenamefont
  {Tobochnik}(1996)}]{GouldHarvey1996Aitc}%
  \BibitemOpen
  \bibfield  {author} {\bibinfo {author} {\bibfnamefont {H.}~\bibnamefont
  {Gould}}\ and\ \bibinfo {author} {\bibfnamefont {J.}~\bibnamefont
  {Tobochnik}},\ }\href@noop {} {\emph {\bibinfo {title} {{An introduction to
  computer simulation methods : applications to physical systems}}}},\ \bibinfo
  {edition} {2nd}\ ed.\ (\bibinfo  {publisher} {Addison-Wesley},\ \bibinfo
  {address} {Reading, Mass.},\ \bibinfo {year} {1996})\BibitemShut {NoStop}%
\bibitem [{\citenamefont {Slattery}\ \emph {et~al.}(1980)\citenamefont
  {Slattery}, \citenamefont {Doolen},\ and\ \citenamefont
  {DeWitt}}]{Slattery:1980}%
  \BibitemOpen
  \bibfield  {author} {\bibinfo {author} {\bibfnamefont {W.~L.}\ \bibnamefont
  {Slattery}}, \bibinfo {author} {\bibfnamefont {G.~D.}\ \bibnamefont
  {Doolen}},\ and\ \bibinfo {author} {\bibfnamefont {H.~E.}\ \bibnamefont
  {DeWitt}},\ }\bibfield  {title} {\bibinfo {title} {{Improved equation of
  state for the classical one-component plasma}},\ }\href
  {https://doi.org/10.1103/PhysRevA.21.2087} {\bibfield  {journal} {\bibinfo
  {journal} {Phys. Rev. A}\ }\textbf {\bibinfo {volume} {21}},\ \bibinfo
  {pages} {2087} (\bibinfo {year} {1980})}\BibitemShut {NoStop}%
\bibitem [{\citenamefont {Stringfellow}\ \emph {et~al.}(1990)\citenamefont
  {Stringfellow}, \citenamefont {DeWitt},\ and\ \citenamefont
  {Slattery}}]{Stringfellow:1990}%
  \BibitemOpen
  \bibfield  {author} {\bibinfo {author} {\bibfnamefont {G.~S.}\ \bibnamefont
  {Stringfellow}}, \bibinfo {author} {\bibfnamefont {H.~E.}\ \bibnamefont
  {DeWitt}},\ and\ \bibinfo {author} {\bibfnamefont {W.~L.}\ \bibnamefont
  {Slattery}},\ }\bibfield  {title} {\bibinfo {title} {{Equation of state of
  the one-component plasma derived from precision Monte Carlo calculations}},\
  }\href {https://doi.org/10.1103/PhysRevA.41.1105} {\bibfield  {journal}
  {\bibinfo  {journal} {Phys. Rev. A}\ }\textbf {\bibinfo {volume} {41}},\
  \bibinfo {pages} {1105} (\bibinfo {year} {1990})}\BibitemShut {NoStop}%
\bibitem [{\citenamefont {Potekhin}\ and\ \citenamefont
  {Chabrier}(2000)}]{Potekhin:2000}%
  \BibitemOpen
  \bibfield  {author} {\bibinfo {author} {\bibfnamefont {A.~Y.}\ \bibnamefont
  {Potekhin}}\ and\ \bibinfo {author} {\bibfnamefont {G.}~\bibnamefont
  {Chabrier}},\ }\bibfield  {title} {\bibinfo {title} {{Equation of state of
  fully ionized electron-ion plasmas. II. Extension to relativistic densities
  and to the solid phase}},\ }\href {https://doi.org/10.1103/PhysRevE.62.8554}
  {\bibfield  {journal} {\bibinfo  {journal} {Phys. Rev. E}\ }\textbf {\bibinfo
  {volume} {62}},\ \bibinfo {pages} {8554} (\bibinfo {year}
  {2000})}\BibitemShut {NoStop}%
\end{thebibliography}

\providecommand{\noopsort}[1]{}\providecommand{\singleletter}[1]{#1}%

\end{document}